\documentclass[%
 reprint,
 amsmath,
 amssymb,
 aps,
 prb,
 superscriptaddress,
 nofootinbib,
]{revtex4-2}

\usepackage{graphicx}
\usepackage[dvipsnames]{xcolor}
\usepackage{dcolumn}
\usepackage{bm}

\usepackage{mathtools}
\usepackage[T1]{fontenc}
\usepackage[utf8]{inputenc}
\usepackage{makerobust}
\usepackage{textcomp}
\usepackage[english]{babel}
\usepackage{lmodern}
\usepackage[babel]{microtype}
\usepackage{setspace}
\DeclareUnicodeCharacter{2013}{--} % breaking at ndash

\usepackage{xcolor}
\usepackage[colorlinks,
			linkcolor=black,
			citecolor=black,
			filecolor=black,
			urlcolor=blue!50!black,
			pdfusetitle]{hyperref}
\hypersetup{
	pdfauthor={test},
	pdfsubject={Solid-State Physics},
	pdfkeywords={test}
 }

\usepackage[capitalize]{cleveref}
\newcommand{\makeauthor}[2]{\newcommand{#1}[1]{{%
  \protect%
  \color{#2}{%
    \bfseries\begingroup%
        \escapechar=-1\edef\x{\endgroup\string#1}\x:%
  }\itshape{} ##1}}%
  \MakeRobustCommand#1}

\makeauthor{\jp}{ForestGreen}
\makeauthor{\as}{Emerald}
\makeauthor{\rv}{blue}
\begin{document}

\title{Finite-Temperature Flat-Band Ferromagnetism in the Kagome Hubbard Model}

\author{Alon Strugatsky}
\author{Jonas Profe}
\author{Roser Valentí}
\affiliation{Institut für Theoretische Physik, Goethe-Universität, 60438 Frankfurt am Main, Germany}

\date{\today}

\begin{abstract}
Kagome metals exhibit a rich interplay of topology, electronic correlations, and lattice dynamics. Recent discoveries of Kagome materials with a flat band near the Fermi level have revealed a variety of correlated electronic phases. However, 
 elucidating their microscopic origin remains challenging, as realistic descriptions require accounting for multiple orbitals and competing interactions on an equal footing.
 To disentangle correlation effects from material-specific details and identify the essential physics of the flat-band regime, we study the single-orbital Kagome–Hubbard model at flat-band fillings using dynamical mean-field theory. We find strong signatures of flat-band ferromagnetism, consistent with exact and mean-field ground-state results. Moreover, we uncover an unconventional quasi-ordered phase in which a partially filled spin-polarized flat band pinned at the Fermi level gives rise to persistent local spin fluctuations down to zero temperature, in striking contrast to the classical behavior expected for a conventional ferromagnet. Our results demonstrate that these anomalous fluctuations are an intrinsic consequence of the flat-band degeneracy and establish a minimal framework for understanding correlation effects in flat-band Kagome systems.
\end{abstract}

\maketitle

\section{Introduction}\label{sec:intro}

The synthesis of AV$_3$Sb$_5$ (A = K, Rb, or Cs) established Kagome metals as an important platform for investigating correlated and topological electronic phenomena~\cite{ortiz2019new,ortiz2020cs,wilson2024v3sb5, kagome_metals_review}. These materials exhibit a rich phase diagram, including charge-density-wave orders and unconventional superconductivity~\cite{PhysRevLett.127.046401, Jiang2021, PhysRevLett.126.247001, christensen2021theory, lin2021complex, ortiz2021superconductivity, 
%Mielke2022, 
Asaba2024}. More recently, a new class of Kagome metals has been discovered in which the geometrically induced flat band lies close to the Fermi level~\cite{kang2020dirac,li2021dirac,sun2022observation,hu2022topological, Lou2024, Liu2024,wang2025spin,xie2025electron, Lee2026, crispino_2025, chatzieleftheriou2026pressure}. The position of this flat band provides a unique platform for exploring the interplay of strong electronic correlations and topology: the flat dispersion enhances the effects of interaction through electron localization, while the nontrivial band wavefunctions give rise to topological phenomena~\cite{Peotta2015, kang2020topological, Wang2025, pollmann2008kinetic, Aoki_2025}. As a result, these systems host a variety of emergent phases, including charge order, unconventional superconductivity, and (flat-band) magnetism~\cite{yin2019negative, PhysRevB.106.115139,FeGe, Liu2024, wang2025spin}.

A minimal description of this broad class of materials is provided by the Kagome–Hubbard model~\cite{Veit_1989,kiesel_2012, wang_2013, Kaufmann_2021,romer2022superconductivity,yu2012chiral, ferrari2022charge}. Even in the absence of interactions, the model exhibits a remarkably rich electronic structure, featuring Dirac cones, van Hove singularities, and a flat band. Owing to this diversity of electronic states, it serves as a versatile framework for understanding emergent phenomena across a wide range of Kagome materials~\cite{yu2012chiral, ferrari2022charge,profe2024kagome, Wang_2026}. Importantly, the flat band of the Kagome lattice originates from geometric frustration rather than from localized atomic orbitals, as in many (f)-electron systems. Consequently, its wavefunctions retain a nontrivial momentum dependence, giving rise to a non-trivial quantum geometry~\cite{Peotta2015, yu2025quantum}.

The flat-band regime poses a significant theoretical challenge, since the quenched kinetic energy restricts the applicability of conventional perturbative approaches. Consequently, this parameter regime has received considerably less attention than other fillings of the Kagome–Hubbard model. Early works established an exact ground state within a specific filling range using graph-theoretical methods~\cite{mielke1991ferromagnetic,mielke1991ferromagnetism,mielke1992exact}. Subsequent mean-field studies extended the ground-state phase diagram and uncovered a broader range of competing quasi-ordered phases~\cite{lin2024complex, vidarte2026filling}. More recently, numerically exact finite-temperature investigations have become available~\cite{correia2026finite,wang2026magnetic}, albeit at larger temperatures. Despite these advances, a unified picture that bridges the gap between ground-state properties and finite temperature signatures beyond mean field is still lacking. 

In this work, we address this gap employing single-site dynamical mean-field theory~\cite{metzner1989correlated, georges1996dynamical} to study the flat-band regime in the Kagome Hubbard model. Our results reproduce the analytically established ground-state behavior and connect smoothly to recent finite-temperature quantum Monte Carlo studies. Furthermore, we uncover distinct spin responses within the predicted magnetically quasi-ordered phase and relate them to the underlying spectral properties.

\vspace{-10pt}
\section{Model and Methods}\label{sec:method}
\begin{figure}
    \centering
    \includegraphics[width=\linewidth]{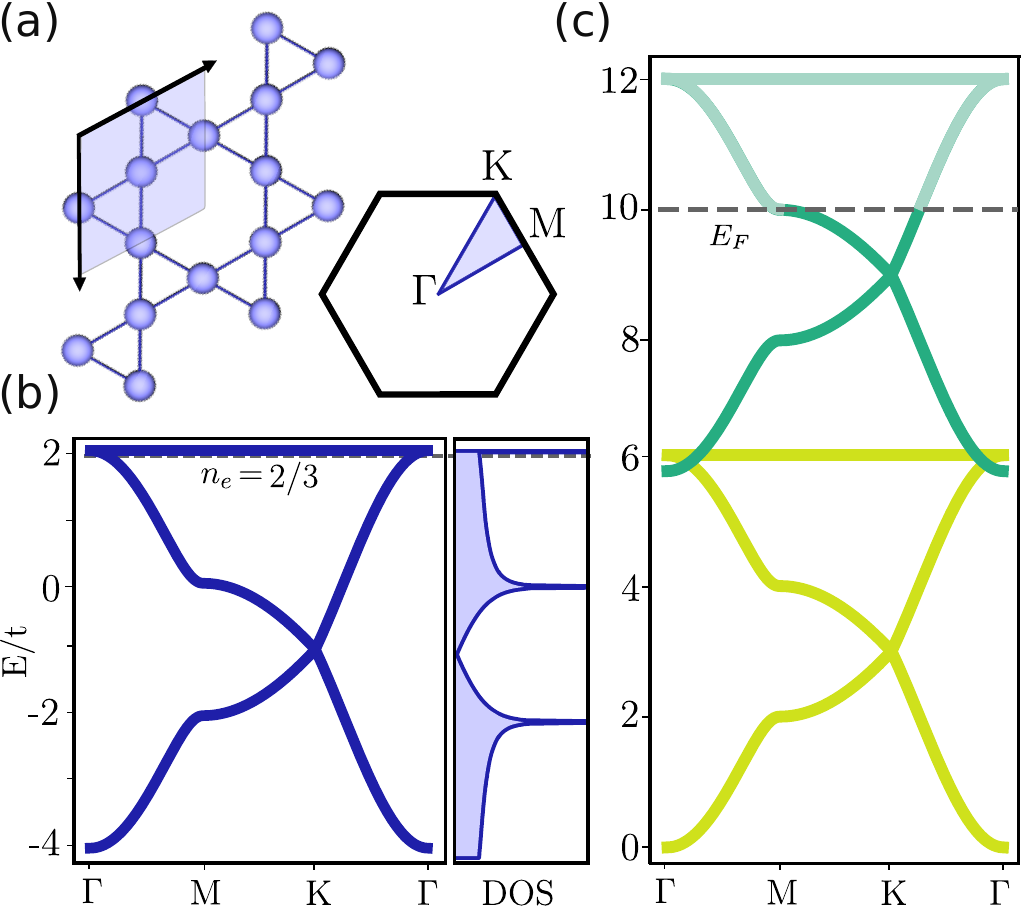}
    % \captionsetup{justification=justified,singlelinecheck=false,font=footnotesize, format=plain, labelformat=simple, labelsep=period, name=FIG.}
    \caption{ (a) The kagome lattice in reciprocal and real space with a single unit cell highlighted by blue shading. (b) The non-interacting bandstructure and density of states. (c) Mean-field spin-split band structure at $n_{FB}=1/8$ into a majority spin band (light green) and a minority spin band (dark green), with the Fermi level crossing the van-Hove singularity of the minority spin band.}
    % \centering
    % \captionsetup{singlelinecheck=off,font=footnotesize}
    \label{fig:non_int_dos}
\end{figure}

We consider the Hubbard model on the kagome lattice,
\begin{equation}
H=-t\sum_{\left< i,j\right>,\sigma}c^{\dagger}_{i\sigma} c_{j\sigma} + U\sum_{i} 
n^{\dagger}_{i\sigma}n_{i\sigma},
\end{equation}
where $c^{\dagger}_{i\sigma}$ ($c_{i\sigma}$) are the electronic creation (annihilation) operators on site $i$ with spin $\sigma$, $\left< i,j\right>$ denotes nearest neighbors, see \cref{fig:non_int_dos}(a), and $n_{i\sigma}=c^{\dagger}_{i\sigma}c_{i\sigma}$ is the density operator.

As a line graph of the bipartite honeycomb lattice, the kagome lattice features two dispersive bands corresponding to bonding and anti-bonding of the honeycomb, as well as an emergent flat band (FB) due to destructive interference, see \cref{fig:non_int_dos}(b). In this work we focus exclusively on partial fillings of this FB. To this end we define the FB filling $n_{FB}=(n_e-4)/2\in[0,1]$, where $n_{FB}=0$  corresponds to an empty flat band  ($2/3$ filling of the entire lattice) and $n_{FB}=1$ to a completely filled system. Further, we define the spin resolved flat band filling as $n^{FB}_{\sigma}=(n_{e;\sigma}-2)$. All quantities are expressed in units of the lattice hopping $t$.

\begin{figure*}        % Use the asterisk here
    \centering
    \includegraphics[width=0.95\textwidth]{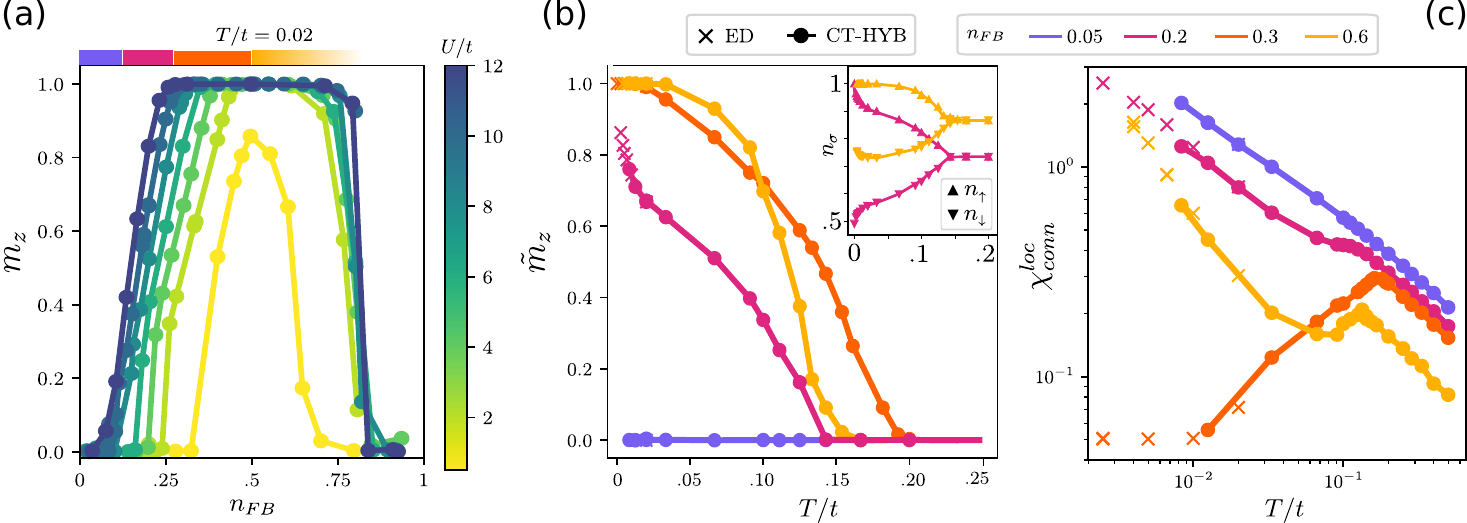}
    \caption{Single-site DMFT results. (a) Evolution of the normalized impurity magnetization as a function of the flat-band filling $n_{FB}$ for different interaction strengths, indicated by the curve colors, at $T/t=0.02$. The four regimes discussed in the text are indicated above the plot: from left to right, the paramagnetic (purple), onset magnetization (magenta), dispersive demagnetization (orange), and flat-band (FB) demagnetization (dark yellow) regimes. (b) Temperature dependence of the normalized magnetization at $U/t=10$ in the four regimes. Saturation is reached only at $T\rightarrow0$, indicating strong competition between the ferromagnetic order and thermal fluctuations. For larger fillings the magnetization saturates already at finite temperatures. The inset shows spin populations. (c) Connected local spin susceptibility from DMFT at $U/t=10$ as a function of temperature. The results reveal three distinct behaviors: a Curie law in the paramagnetic regime, suppression of spin fluctuations in the conventional ferromagnetic regime, and a persistent Curie-like response in the flat-band quasi-ordered regimes, as discussed in the text.
    }
    \label{fig:mag_figs}
\end{figure*}

In order to study the effect of electronic correlations, we employ dynamical mean-field theory (DMFT)~\cite{metzner1989correlated, PhysRevB.45.6479, georges1996dynamical}. Within DMFT, the lattice Hubbard model is mapped onto a quantum impurity problem embedded in a self-consistent bath. Since the non-interacting  Hamiltonian is a 3$\times$3 matrix in the sub-lattice basis, the impurity problem is in principle a 3-site cluster with a 3$\times$3 self energy $\Sigma^{\rm imp}_{ij}(i\omega_n)$ and hybridization $\Delta_{ij}(i\omega_n)$,
whose off-diagonal elements encode inter-sub-lattice correlations. Here we simplify this cluster 
retaining only the diagonal contributions of $\Sigma_{\rm imp}$ and the Green's function $G_{\rm imp}$. Thus, the self consistency loop is closed by
\begin{equation}
G_{\rm loc}(i\omega_n) = \frac{1}{3N_{\mathbf k}}\sum_{\mathbf k,\alpha}\Big[(i\omega_n+\mu)\mathbb{I}-H^0_{ij}(\mathbf{k})-\Sigma(i\omega_n)\mathbb{I}\Big]^{-1}_{\alpha\alpha},
\end{equation}
where $H^{0}_{ij}(\mathbf{k})$ is the non-interacting Hamiltonian. The diagonal elements are related through the lattice $C_3$ symmetry, such that we need to solve a single Anderson impurity problem at each iteration. Due to enforcing this symmetry, we cannot capture charge orderings. 
To solve the effective one-site impurity model we use the continuous time hybridization expansion (CT-HYB) Quantum Monte Carlo (QMC) solver \texttt{W2dynamics}~\cite{gull2011continuous, wallerberger2019w2dynamics}, as well as the exact diagonalization (ED) solver \texttt{EDIpack}~\cite{amaricci2022CPC,Crippa2025SPC,Crippa2025SPCa} in order to access the low temperature regime. We find good agreement between results from CT-HYB and EDIpack at low temperatures. 
In order to access the real-frequency spectral properties of the model, the results on the Matsubara axis are analytically continued using the maximum entropy method~\cite{jarrell1996maximum} utilizing the \texttt{ana\_cont}~\cite{kaufmann2023ana_cont} library.

\section{Results and discussion}\label{intro}

In the following, we discuss the magnetic phase diagram obtained within single-site DMFT. The finite-temperature ordered states found in our simulations should not be interpreted as evidence for a true long-range magnetic order in the strictly two-dimensional system, where long-wavelength fluctuations suppress such order at finite temperature~\cite{PhysRevLett.17.1133}. Rather, the finite-temperature transitions found within DMFT are indicative of a regime of strong ferromagnetic correlations and a growing magnetic correlation length. Given the known ferromagnetic nature of the zero-temperature ground state~\cite{mielke1992exact}, we therefore interpret these transitions as signatures of a precursor regime to the zero-temperature ordered state, characterized by pronounced ferromagnetic correlations.

\begin{figure*}        
    \centering
    \includegraphics[width=0.97\textwidth]{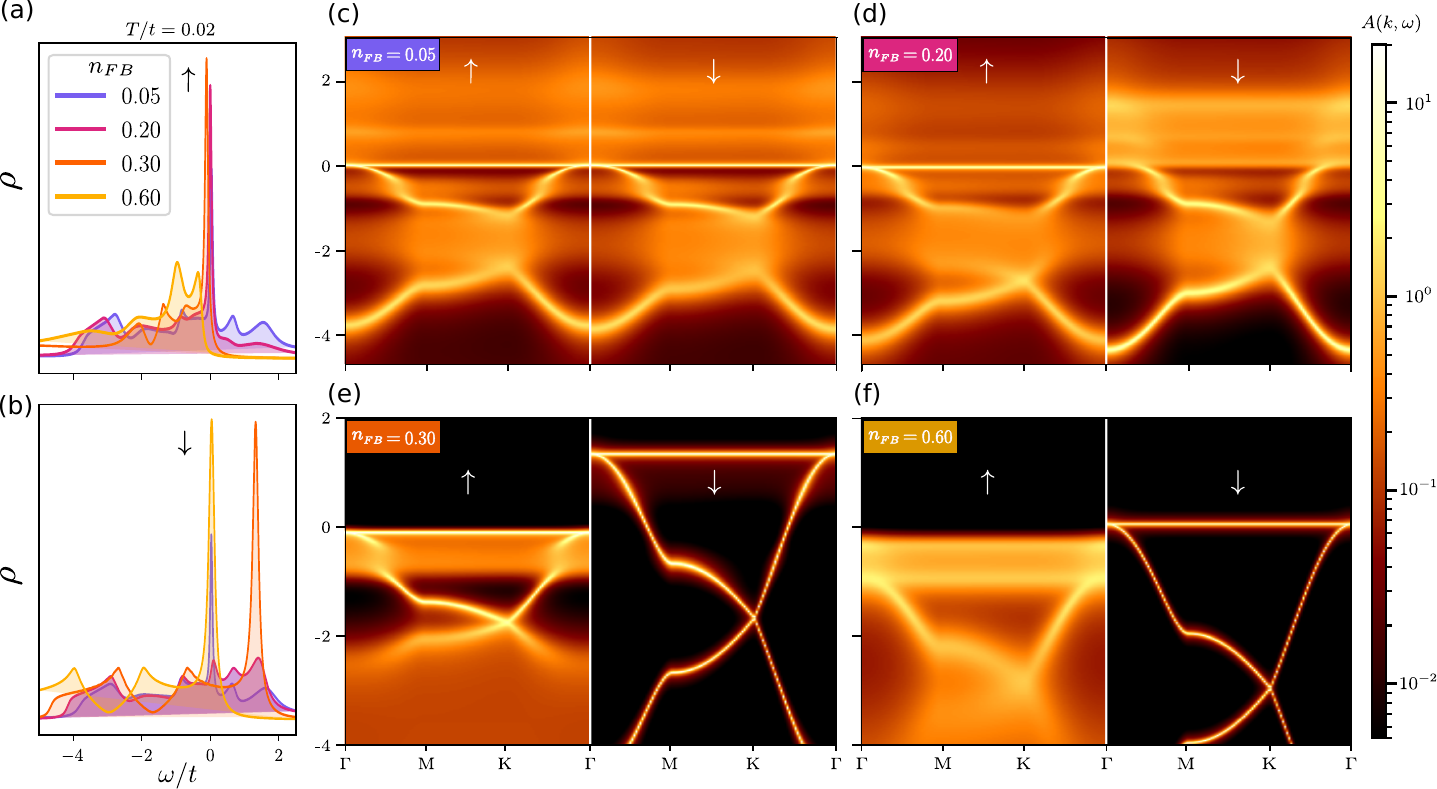}
    \caption{ Spin-resolved local density of states (DOS) for the majority (a) and minority (b) spin channels, together with the momentum-resolved spectral functions (c)-(f) obtained from DMFT. The Fermi level is set to $E_F=0$. In the onset regime, the minority-spin channel exhibits a strong loss of quasiparticle weight near the Fermi level (\cref{fig:Akw_Gw}(d)). In contrast, in the saturated regime, the minority-spin quasiparticles remain fully coherent ((e,f)).
    }
    \label{fig:Akw_Gw}
\end{figure*}

 \cref{fig:mag_figs}(a) shows the low temperature magnetization behavior at different flat-band (FB) fillings. To visualize whether the magnetization is saturated, we define the normalized magnetization
  $\tilde{m}_z \equiv (n^{FB}_{\uparrow} - n^{FB}_{\downarrow})/(2 - (n^{FB}_{\uparrow} + n^{FB}_{\downarrow})$, which equals unity for a fully saturated magnetization at any partial filling of the flat band and vanishes in the paramagnetic state. 
  We observe four distinct magnetic regimes which are marked by colors in \cref{fig:mag_figs}(a). First, at low fillings, a paramagnetic regime persists up to a critical filling of $n^c_{FB}\approx1/8$, marked in purple in \cref{fig:mag_figs}(a). 
  Second, starting at $n^c_{FB}\approx1/8$, we enter the onset regime of magnetization, marked in magenta in \cref{fig:mag_figs}(a). In this regime, DMFT predicts a finite magnetization, although it remains below its saturated value at finite temperatures. The critical filling for the transition, $n^c_{FB}\approx1/8$, corresponds to the filling at which the upper van Hove singularity of the minority band (\cref{fig:non_int_dos} (c)) is crossing the Fermi-level in a mean-field picture. Third, we enter the {dispersive}
demagnetization regime, marked in orange in \cref{fig:mag_figs}(a). This regime is characterized by saturated magnetization and progressive filling of the dispersive minority-spin band, see \cref{fig:Akw_Gw}(e). Lastly, we enter the flat-band demagnetization regime, marked in dark yellow in \cref{fig:mag_figs}(a). In this regime the minority-spin flat band is populated, see \cref{fig:Akw_Gw}(f). We will discuss below why we distinguish these two demagnetization regimes.
Finally, we observe a breakdown of the finite magnetization at approximately  $n_{FB}=3/4$, consistent with previous studies~\cite{mielke1992exact}.

The temperature dependence of the magnetization in the strong-coupling limit ($U/t=10$) is shown in \cref{fig:mag_figs}(b), with one representative filling for each regime. For each of these regimes, we identify the critical temperature $T_c$ as the temperature at which we observe a finite magnetization. Further, we define the saturation temperature $T_s$ as the temperature at which the magnetization reaches approximately unity. Both of the demagnetization regimes display a finite saturation temperature $T_s$ indicating a mean-field like ferromagnetic state. In contrast, in the onset regime (magenta data in \cref{fig:mag_figs}(b)) $T_s=0$ and $\tilde{m}(T)$ displays a non-monotonic behavior with an inflection point. Below this inflection point, magnetization rises rapidly, which is caused by a rapid depopulation of the minority spin, as shown in the inset of \cref{fig:mag_figs}(b). This behavior originates from the spectral weight redistribution as the (minority spin) flat band is pushed above the Fermi level by the increasing magnetization, as displayed in \cref{fig:Akw_Gw}(d).

To investigate the interplay between magnetic order and fluctuations, we analyze the local spin susceptibility. Within DMFT, the local spin susceptibility $\chi_{\mathrm{loc}}$ can be computed directly from the impurity model. Since the calculations are performed in the symmetry-broken phase, we consider its connected part, $\chi^{\mathrm{conn}}_{\mathrm{loc}}$, which removes the trivial Curie-like $1/T$ contribution arising from the finite ordered moment, $\langle S_z\rangle\neq0$,
\begin{equation}
\chi^{\text{conn}}_{\text{loc}} = \intop_{0}^{\beta}d\tau\left(\,\left<S_{z}(\tau)S_z(0)\right>\,\right)\,\,-\,\beta\left<S_z\right>^2.
\end{equation}
Thereby, we isolate the dynamical spin fluctuations beyond the static ordered moment. The resulting temperature dependence of $\chi^{\mathrm{conn}}_{\mathrm{loc}}$ at strong coupling ($U/t=10$) is shown in \cref{fig:mag_figs}(c). In the paramagnetic regime (purple curve), $\chi^{\mathrm{conn}}_{\mathrm{loc}}$ follows a Curie law, $\chi^{\mathrm{conn}}_{\mathrm{loc}}\sim 1/T$, which is consistent with the presence of free, unscreened local moments. In the dispersive demagnetization regime (orange curve), $\chi^{\mathrm{conn}}_{\mathrm{loc}}$ drops sharply below $T_c$ and vanishes below $T_s$, identifying this phase as a true mean-field like ferromagnet in which spin fluctuations are completely frozen out. By contrast, substantial residual spin fluctuations persist down to $T\rightarrow0$ in both the onset (magenta curve) and the FB demagnetization regimes (yellow curve).

To understand the origin of the anomalous behavior, we consider the momentum-resolved spectral function $A(k,\omega)$ (\cref{fig:Akw_Gw}). In the paramagnetic regime, see \cref{fig:Akw_Gw}(c), we find satellite peaks in the spectral function extending up to $2$t above the Fermi-level, see also the density of states for majority and minority spin species (\cref{fig:Akw_Gw}(a) and~(b)). Further we observe a strong incoherence of the dispersive bands at around $-0.5$t. Remarkably, the flat band itself remains relatively sharp. 
In the onset regime (\cref{fig:Akw_Gw}(d)), the minority-spin spectral function is strongly broadened near the Fermi level, indicating a significant reduction of quasiparticle weight. Moreover, the satellites above the Fermi level carry more spectral weight than the flat band itself, see \cref{fig:Akw_Gw}(a) and (b). In contrast, the majority-spin channel becomes more coherent.
 Due to the strong incoherence of the minority spin, the ordered moment remains susceptible to thermal fluctuations even below $T_c$. This prevents complete saturation except in the limit $T\rightarrow0$ and thereby explaining the anomalous behavior of $\tilde{m}_z(T)$ (\cref{fig:mag_figs} (b)). In the dispersive demagnetization regime (\cref{fig:Akw_Gw}(e)), the majority-spin band is fully occupied, while the minority-spin band is filled only up to a dispersive band. Consequently, the minority-spin quasiparticles are essentially non-interacting and remain coherent, i.e.
 ~$\Im\Sigma(\omega)\rightarrow0$. This is consistent with the mean-field like behavior of both $\tilde{m}_z(T)$ and $\chi^{\mathrm{conn}}_{\mathrm{loc}}(T)$. We find a similar picture in the flat-band demagnetization regime as well, see \cref{fig:Akw_Gw}(f), with the crucial difference that the minority spin flat-band is pinned to the Fermi-level instead of a dispersive state.

From these results, we can understand the persistent Curie-behavior of the local spin-susceptibility we observed in \cref{fig:mag_figs}(c): In both the onset and the FB demagnetization regimes the Fermi level resides within a flat band of one spin species, see \cref{fig:Akw_Gw} (d) and (f). The presence of this flat band at the Fermi-level, pinned by the filling constraints, causes a divergent density of states which in turn causes the observed Curie-behavior. The additional broadening of the FB we have in the onset regime alters the slope of the local susceptibility. 
We thus identify a flat band at the Fermi level as the key ingredient underlying the anomalous persistence of local spin fluctuations deep inside the quasi-ordered state. We stress that this is a key point for future studies as such persistent spin-fluctuations in the ordered state could destroy magnetic orders even in a higher dimensional system. Such a situation could be realized in the pyrochlore lattice~\cite{PhysRevLett.112.207202}.

Since a flat band at the Fermi-level is the central ingredient to the observed behavior, one could assume that the quantum geometry of the Bloch states is central for its appearance~\cite{kitamura2024spin,kitamura2026quantum}. However, within single-site DMFT the effective impurity bath is determined solely by the local non-interacting density of states through the self-consistency condition, rendering the local impurity problem insensitive to the spatial structure and topology of the Bloch wave functions. We therefore conclude that the persistent low-temperature divergence of $\chi_{\mathrm{loc}}^{\mathrm{conn}}$ is not driven by the quantum geometry of the Bloch states, but instead truly originates from the macroscopic degeneracy associated with the flat band itself. %This interpretation is consistent with the fact that the divergence appears whenever the Fermi level resides within a flat band, independently of its quantum geometry.

In summary, we have studied flat-band ferromagnetism in the single-orbital Kagome-Hubbard model employing dynamical mean-field theory. We find a ferromagnetic order consistent with exact and mean-field predictions and identify an unconventional quasi-ordered regime characterized by persistent local spin fluctuations down to zero temperature. These fluctuations are associated with a spin-polarized flat band pinned at the Fermi level, highlighting the distinctive role of flat-band degeneracy in shaping magnetic correlations. Our DMFT results further reveal characteristic temperature scales for the development of strong ferromagnetic correlations, providing insight into the precursor regime of the ferromagnetic ground state. %More broadly, the interplay between magnetic order, local fluctuations, and (quasi-)flat bands at the Fermi level offers a rich setting for correlated physics and provides valuable insights into emerging flat-band Kagome materials~\cite{kang2020dirac,li2021dirac,sun2022observation,hu2022topological, Lou2024, Liu2024,Lee2026}.
\begin{acknowledgments}
We would like to thank Lorenzo Crippa for useful discussions at the various stages of the work. We acknowledge support by the Deutsche Forschungsgemeinschaft (DFG, German Research Foundation) for funding through projects QUAST-FOR5249 (449872909) (projects P4). The authors gratefully acknowledge the computing time provided to them at the NHR Center NHR@SW at Goethe-University Frankfurt. This is funded by the Federal Ministry of Education and Research, and the state governments participating on the basis of the resolutions of the GWK for national high performance computing at universities (www.nhr-verein.de/unsere-partner).
\end{acknowledgments}

\bibliography{references}

@article{ortiz2019new,
  title={New kagome prototype materials: discovery of {KV}$_3${S}b$_5$, {R}b{V}$_3${S}b$_5$, and {C}s{V}$_3${S}b$_5$},
  author={Ortiz, Brenden R and Gomes, L{\'\i}dia C and Morey, Jennifer R and Winiarski, Michal and Bordelon, Mitchell and Mangum, John S and Oswald, Iain W and Rodriguez-Rivera, Jose A and Neilson, James R and Wilson, Stephen D and others},
  journal={Phys. Rev. Mater.},
  volume={3},
  number={9},
  year={2019},
  url={https://doi.org/10.1103/PhysRevMaterials.3.094407}
}

@article{wang2025spin,
  title={Spin excitations and flat electronic bands in a Cr-based kagome superconductor},
  author={Wang, Zehao and Guo, Yucheng and Huang, Hsiao-Yu and Xie, Fang and Huang, Yuefei and Gao, Bin and Oh, Ji Seop and Wu, Han and Okamoto, Jun and Channagowdra, Ganesha and others},
  journal={Nature Communications},
  volume={16},
  number={1},
  pages={7573},
  year={2025},
  publisher={Nature Publishing Group UK London}
}

@article{xie2025electron,
  title={Electron correlations in the kagome flat band metal CsCr 3 Sb 5},
  author={Xie, Fang and Fang, Yuan and Li, Ying and Huang, Yuefei and Chen, Lei and Setty, Chandan and Sur, Shouvik and Yakobson, Boris and Valent{\'\i}, Roser and Si, Qimiao},
  journal={Physical Review Research},
  volume={7},
  number={2},
  pages={L022061},
  year={2025},
  publisher={APS}
}

@article{PhysRevLett.112.207202,
  title = {Hubbard Model on the Pyrochlore Lattice: A 3D Quantum Spin Liquid},
  author = {Normand, B. and Nussinov, Z.},
  journal = {Phys. Rev. Lett.},
  volume = {112},
  issue = {20},
  pages = {207202},
  numpages = {5},
  year = {2014},
  month = {May},
  publisher = {American Physical Society},
  doi = {10.1103/PhysRevLett.112.207202},
  url = {https://link.aps.org/doi/10.1103/PhysRevLett.112.207202}
}

@article{Liu2024,
  title = {Superconductivity under pressure in a chromium-based kagome metal},
  volume = {632},
  ISSN = {1476-4687},
  url = {http://dx.doi.org/10.1038/s41586-024-07761-x},
  DOI = {10.1038/s41586-024-07761-x},
  number = {8027},
  journal = {Nature},
  publisher = {Springer Science and Business Media LLC},
  author = {Liu,  Yi and Liu,  Zi-Yi and Bao,  Jin-Ke and Yang,  Peng-Tao and Ji,  Liang-Wen and Wu,  Si-Qi and Shen,  Qin-Xin and Luo,  Jun and Yang,  Jie and Liu,  Ji-Yong and Xu,  Chen-Chao and Yang,  Wu-Zhang and Chai,  Wan-Li and Lu,  Jia-Yi and Liu,  Chang-Chao and Wang,  Bo-Sen and Jiang,  Hao and Tao,  Qian and Ren,  Zhi and Xu,  Xiao-Feng and Cao,  Chao and Xu,  Zhu-An and Zhou,  Rui and Cheng,  Jin-Guang and Cao,  Guang-Han},
  year = {2024},
  month = Aug,
  pages = {1032–1037}
}

@misc{crispino_2025,
      title={Tunable Electronic Correlations in 135-Kagome Metals}, 
      author={Matteo Crispino and Niklas Witt and Stefan Enzner and Tommaso Gorni and Luca de' Medici and Domenico Di Sante and Giorgio Sangiovanni},
      year={2025},
      eprint={2512.22576},
      archivePrefix={arXiv},
      url={https://arxiv.org/abs/2512.22576}, 
}

@article{Veit_1989,
  title = {Nuclear antiferromagnetism in a registered $^{3}\mathrm{He}$ solid},
  author = {Elser, Veit},
  journal = {Phys. Rev. Lett.},
  volume = {62},
  issue = {20},
  pages = {2405--2408},
  numpages = {0},
  year = {1989},
  month = {May},
  publisher = {American Physical Society},
  doi = {10.1103/PhysRevLett.62.2405},
  url = {https://link.aps.org/doi/10.1103/PhysRevLett.62.2405}
}

@article{kiesel_2012,
  title = {Sublattice interference in the kagome Hubbard model},
  author = {Kiesel, Maximilian L. and Thomale, Ronny},
  journal = {Phys. Rev. B},
  volume = {86},
  issue = {12},
  pages = {121105(R)},
  numpages = {4},
  year = {2012},
  month = {Sep},
  publisher = {American Physical Society},
  doi = {10.1103/PhysRevB.86.121105},
  url = {https://link.aps.org/doi/10.1103/PhysRevB.86.121105}
}

@article{wang_2013,
  title = {Competing electronic orders on kagome lattices at van Hove filling},
  author = {Wang, Wan-Sheng and Li, Zheng-Zhao and Xiang, Yuan-Yuan and Wang, Qiang-Hua},
  journal = {Phys. Rev. B},
  volume = {87},
  issue = {11},
  pages = {115135},
  numpages = {8},
  year = {2013},
  month = {Mar},
  publisher = {American Physical Society},
  doi = {10.1103/PhysRevB.87.115135},
  url = {https://link.aps.org/doi/10.1103/PhysRevB.87.115135}
}

@article{ortiz2020cs,
  title={{C}s{V}$_3${S}b$_5$: a $\mathbb{Z}_2$ topological kagome metal with a superconducting ground state},
  author={Ortiz, Brenden R and Teicher, Samuel ML and Hu, Yong and Zuo, Julia L and Sarte, Paul M and Schueller, Emily C and Abeykoon, AM Milinda and Krogstad, Matthew J and Rosenkranz, Stephan and Osborn, Raymond and others},
  journal={Phys. Rev. Lett.},
  volume={125},
  number={24},
  pages={247002},
  year={2020},
  publisher={APS},
  url={https://doi.org/10.1103/PhysRevLett.125.247002}
}

@article{FeGe,
  title = {Discovery of Charge Order and Corresponding Edge State in Kagome Magnet {F}e{G}e},
  author = {Yin, Jia-Xin and Jiang, Yu-Xiao and Teng, Xiaokun and Hossain, Md. Shafayat and Mardanya, Sougata and Chang, Tay-Rong and Ye, Zijin and Xu, Gang and Denner, M. Michael and Neupert, Titus and Lienhard, Benjamin and Deng, Han-Bin and Setty, Chandan and Si, Qimiao and Chang, Guoqing and Guguchia, Zurab and Gao, Bin and Shumiya, Nana and Zhang, Qi and Cochran, Tyler A. and Multer, Daniel and Yi, Ming and Dai, Pengcheng and Hasan, M. Zahid},
  journal = {Phys. Rev. Lett.},
  volume = {129},
  issue = {16},
  pages = {166401},
  numpages = {7},
  year = {2022},
  month = {Oct},
  publisher = {American Physical Society},
  doi = {10.1103/PhysRevLett.129.166401},
  url = {https://link.aps.org/doi/10.1103/PhysRevLett.129.166401}
}

@article{PhysRevB.106.115139,
  title = {Uniaxial ferromagnetism in the kagome metal {T}b{V}$_6${S}n$_6$},
  author = {Rosenberg, Elliott and DeStefano, Jonathan M. and Guo, Yucheng and Oh, Ji Seop and Hashimoto, Makoto and Lu, Donghui and Birgeneau, Robert J. and Lee, Yongbin and Ke, Liqin and Yi, Ming and Chu, Jiun-Haw},
  journal = {Phys. Rev. B},
  volume = {106},
  issue = {11},
  pages = {115139},
  numpages = {8},
  year = {2022},
  month = {Sep},
  publisher = {American Physical Society},
  doi = {10.1103/PhysRevB.106.115139},
  url = {https://link.aps.org/doi/10.1103/PhysRevB.106.115139}
}

@article{christensen2021theory,
  title={Theory of the charge density wave in {AV}$_3${S}b$_5$ kagome metals},
  author={Christensen, Morten H and Birol, Turan and Andersen, Brian M and Fernandes, Rafael M},
  journal={Phys. Rev. B},
  volume={104},
  number={21},
  pages={214513},
  year={2021},
  publisher={APS},
  url={https://doi.org/10.1103/PhysRevB.104.214513}
}

@article{lin2021complex,
  title={Complex charge density waves at Van Hove singularity on hexagonal lattices: Haldane-model phase diagram and potential realization in the kagome metals {AV}$_3${S}b$_5$ ({A} = {K}, {R}b, {C}s)},
  author={Lin, Yu-Ping and Nandkishore, Rahul M},
  journal={Phys. Rev. B},
  volume={104},
  number={4},
  pages={045122},
  year={2021},
  publisher={APS},
  url={https://doi.org/10.1103/PhysRevB.104.045122}
}

@article{ortiz2021superconductivity,
  title={Superconductivity in the $\mathbb{Z}_2$ kagome metal {KV}$_3${S}b$_5$},
  author={Ortiz, Brenden R and Sarte, Paul M and Kenney, Eric M and Graf, Michael J and Teicher, Samuel ML and Seshadri, Ram and Wilson, Stephen D},
  journal={Phys. Rev. Mater.},
  volume={5},
  number={3},
  pages={034801},
  year={2021},
  publisher={APS},
  url={https://doi.org/10.1103/PhysRevMaterials.5.034801}
}

@article{wilson2024v3sb5,
  title={{AV}$_3${S}b$_5$ kagome superconductors},
  author={Wilson, Stephen D and Ortiz, Brenden R},
  journal={Nat. Rev. Mater.},
  volume={9},
  number={6},
  pages={420--432},
  year={2024},
  publisher={Nature Publishing Group UK London},
  url={https://doi.org/10.1038/s41578-024-00677-y}
}

@article{PhysRevLett.126.247001,
  title = {Double Superconducting Dome and Triple Enhancement of ${T}_{c}$ in the Kagome Superconductor {C}s{V}$_3${S}b$_5$ under High Pressure},
  author = {Chen, K. Y. and Wang, N. N. and Yin, Q. W. and Gu, Y. H. and Jiang, K. and Tu, Z. J. and Gong, C. S. and Uwatoko, Y. and Sun, J. P. and Lei, H. C. and Hu, J. P. and Cheng, J.-G.},
  journal = {Phys. Rev. Lett.},
  volume = {126},
  issue = {24},
  pages = {247001},
  numpages = {8},
  year = {2021},
  month = {Jun},
  publisher = {American Physical Society},
  doi = {10.1103/PhysRevLett.126.247001},
  url = {https://link.aps.org/doi/10.1103/PhysRevLett.126.247001}
}

@article{Jiang2021,
  title = {Unconventional chiral charge order in kagome superconductor {KV}$_3${S}b$_5$},
  volume = {20},
  ISSN = {1476-4660},
  url = {http://dx.doi.org/10.1038/s41563-021-01034-y},
  DOI = {10.1038/s41563-021-01034-y},
  number = {10},
  journal = {Nat. Mater.},
  publisher = {Springer Science and Business Media LLC},
  author = {Jiang,  Yu-Xiao and Yin,  Jia-Xin and Denner,  M. Michael and Shumiya,  Nana and Ortiz,  Brenden R. and Xu,  Gang and Guguchia,  Zurab and He,  Junyi and Hossain,  Md Shafayat and Liu,  Xiaoxiong and Ruff,  Jacob and Kautzsch,  Linus and Zhang,  Songtian S. and Chang,  Guoqing and Belopolski,  Ilya and Zhang,  Qi and Cochran,  Tyler A. and Multer,  Daniel and Litskevich,  Maksim and Cheng,  Zi-Jia and Yang,  Xian P. and Wang,  Ziqiang and Thomale,  Ronny and Neupert,  Titus and Wilson,  Stephen D. and Hasan,  M. Zahid},
  year = {2021},
  pages = {1353–1357},
  url={https://doi.org/10.1038/s41563-021-01034-y}
}

@article{PhysRevLett.127.046401,
  title = {Charge Density Waves and Electronic Properties of Superconducting Kagome Metals},
  author = {Tan, Hengxin and Liu, Yizhou and Wang, Ziqiang and Yan, Binghai},
  journal = {Phys. Rev. Lett.},
  volume = {127},
  issue = {4},
  pages = {046401},
  numpages = {6},
  year = {2021},
  month = {Jul},
  publisher = {American Physical Society},
  doi = {10.1103/PhysRevLett.127.046401},
  url = {https://link.aps.org/doi/10.1103/PhysRevLett.127.046401}
}

@article{hu2022topological,
  title={Topological surface states and flat bands in the kagome superconductor {C}s{V}$_3${S}b$_5$},
  author={Hu, Yong and Teicher, Samuel ML and Ortiz, Brenden R and Luo, Yang and Peng, Shuting and Huai, Linwei and Ma, Junzhang and Plumb, Nicholas C and Wilson, Stephen D and He, Junfeng and others},
  journal={Science Bulletin},
  volume={67},
  number={5},
  pages={495--500},
  year={2022},
  publisher={Elsevier},
  url={https://doi.org/10.1016/j.scib.2021.11.026}
}

@article{li2021dirac,
  title={Dirac cone, flat band and saddle point in kagome magnet {YM}n$_6${S}n$_6$},
  author={Li, Man and Wang, Qi and Wang, Guangwei and Yuan, Zhihong and Song, Wenhua and Lou, Rui and Liu, Zhengtai and Huang, Yaobo and Liu, Zhonghao and Lei, Hechang and others},
  journal={Nat. Commun.},
  volume={12},
  number={1},
  pages={3129},
  year={2021},
  publisher={Nature Publishing Group UK London},
  url={https://doi.org/10.1038/s41467-021-23536-8}
}

@article{Peotta2015,
  title = {Superfluidity in topologically nontrivial flat bands},
  volume = {6},
  ISSN = {2041-1723},
number = {1},
  journal = {Nat. Commun.},
  publisher = {Springer Science and Business Media LLC},
  author = {Peotta,  Sebastiano and T\"{o}rm\"{a},  P\"{a}ivi},
  year = {2015},
  month = Nov,
  url = {http://dx.doi.org/10.1038/ncomms9944}
}

@article{Aoki_2025,
author = {Hideo Aoki},
title = {Flat bands in condensed-matter systems – perspective for magnetism and superconductivity},
journal = {Contemp. Phys.},
volume = {66},
number = {1-4},
pages = {1--38},
year = {2025},
publisher = {Taylor \& Francis},

URL = {https://doi.org/10.1080/00107514.2025.2550105}

}

@article{Wang2025,
  title = {Intriguing kagome topological materials},
  volume = {10},
  ISSN = {2397-4648},
  number = {1},
  journal = {npj Quantum Mater.},
  publisher = {Springer Science and Business Media LLC},
  author = {Wang,  Qi and Lei,  Hechang and Qi,  Yanpeng and Felser,  Claudia},
  year = {2025},
  month = July,
  url = {http://dx.doi.org/10.1038/s41535-025-00790-3},
}

@article{Asaba2024,
  title = {Evidence for an odd-parity nematic phase above the charge-density-wave transition in a kagome metal},
  volume = {20},
  ISSN = {1745-2481},
  url = {http://dx.doi.org/10.1038/s41567-023-02272-4},
  DOI = {10.1038/s41567-023-02272-4},
  number = {1},
  journal = {Nat. Phys.},
  publisher = {Springer Science and Business Media LLC},
  author = {Asaba,  T. and Onishi,  A. and Kageyama,  Y. and Kiyosue,  T. and Ohtsuka,  K. and Suetsugu,  S. and Kohsaka,  Y. and Gaggl,  T. and Kasahara,  Y. and Murayama,  H. and Hashimoto,  K. and Tazai,  R. and Kontani,  H. and Ortiz,  B. R. and Wilson,  S. D. and Li,  Q. and Wen,  H. -H. and Shibauchi,  T. and Matsuda,  Y.},
  year = {2024},
  month = Jan,
  pages = {40–46}
}

@article{PhysRevB.45.6479,
  title = {Hubbard model in infinite dimensions},
  author = {Georges, Antoine and Kotliar, Gabriel},
  journal = {Phys. Rev. B},
  volume = {45},
  issue = {12},
  pages = {6479--6483},
  numpages = {0},
  year = {1992},
  month = {Mar},
  publisher = {American Physical Society},
  doi = {10.1103/PhysRevB.45.6479},
  url = {https://link.aps.org/doi/10.1103/PhysRevB.45.6479}
}

@article{Lee2026,
  title = {Coexisting kagome and heavy fermion flat bands in {Y}b{C}r$_6${G}e$_6$},
  volume = {17},
  ISSN = {2041-1723},
  number = {1},
  journal = {Nat. Commun.},
  publisher = {Springer Science and Business Media LLC},
  author = {Lee,  Hanoh and Lyi,  Churlhi and Lee,  Taehee and Na,  Hyeonhui and Kim,  Jinyoung and Lee,  Sangjae and Kim,  Younsik and Rajapitamahuni,  Anil and Kundu,  Asish K. and Vescovo,  Elio and Park,  Byeong-Gyu and Kim,  Changyoung and Ahn,  Charles H. and Walker,  Frederick J. and Oh,  Ji Seop and Jang,  Bo Gyu and Kim,  Youngkuk and Sohn,  Byungmin and Park,  Tuson},
  year = {2026},
  month = Mar,
  url = {http://dx.doi.org/10.1038/s41467-026-70958-3}
}

@article{Lou2024,
  title = {Orbital-selective effect of spin reorientation on the Dirac fermions in a non-charge-ordered kagome ferromagnet {F}e$_3${G}e},
  volume = {15},
  ISSN = {2041-1723},
  number = {1},
  journal = {Nat. Commun.},
  publisher = {Springer Science and Business Media LLC},
  author = {Lou,  Rui and Zhou,  Liqin and Song,  Wenhua and Fedorov,  Alexander and Tu,  Zhijun and Jiang,  Bei and Wang,  Qi and Li,  Man and Liu,  Zhonghao and Chen,  Xuezhi and Rader,  Oliver and B\"{u}chner,  Bernd and Sun,  Yujie and Weng,  Hongming and Lei,  Hechang and Wang,  Shancai},
  year = {2024},
  month = Nov,
  url = {http://dx.doi.org/10.1038/s41467-024-53343-w}
}

@article{chatzieleftheriou2026pressure,
  title={Pressure Tuning of Electronic Correlations and Flat Bands in {C}s{C}r$_3${S}b$_5$},
  author={Chatzieleftheriou, Maria and Profe, Jonas B and Li, Ying and Valent{\'\i}, Roser},
  journal={arXiv preprint arXiv:2601.14439},
  year={2026},
  url={https://arxiv.org/abs/2601.14439}
}

@article{kang2020dirac,
  title={Dirac fermions and flat bands in the ideal kagome metal {F}e{S}n},
  author={Kang, Mingu and Ye, Linda and Fang, Shiang and You, Jhih-Shih and Levitan, Abe and Han, Minyong and Facio, Jorge I and Jozwiak, Chris and Bostwick, Aaron and Rotenberg, Eli and others},
  journal={Nat. Mater.},
  volume={19},
  number={2},
  pages={163--169},
  year={2020},
  publisher={Nature Publishing Group UK London},
  url={https://doi.org/10.1038/s41563-019-0531-0}
}

@article{kang2020topological,
  title={Topological flat bands in frustrated kagome lattice {C}o{S}n},
  author={Kang, Mingu and Fang, Shiang and Ye, Linda and Po, Hoi Chun and Denlinger, Jonathan and Jozwiak, Chris and Bostwick, Aaron and Rotenberg, Eli and Kaxiras, Efthimios and Checkelsky, Joseph G and others},
  journal={Nat. Commun.},
  volume={11},
  number={1},
  pages={4004},
  year={2020},
  publisher={Nature Publishing Group UK London},
  url={https://doi.org/10.1038/s41467-020-17465-1}
}

@article{sun2022observation,
  title={Observation of topological flat bands in the kagome semiconductor {N}b$_3${C}l$_8$},
  author={Sun, Zhenyu and Zhou, Hui and Wang, Cuixiang and Kumar, Shiv and Geng, Daiyu and Yue, Shaosheng and Han, Xin and Haraguchi, Yuya and Shimada, Kenya and Cheng, Peng and others},
  journal={Nano Letters},
  volume={22},
  number={11},
  pages={4596--4602},
  year={2022},
  publisher={ACS Publications},
  url={https://doi.org/10.1021/acs.nanolett.2c00778}
}

@article{pollmann2008kinetic,
  title={Kinetic ferromagnetism on a kagome lattice},
  author={Pollmann, F and Fulde, P and Shtengel, K},
  journal={Phys. Rev. Lett.},
  volume={100},
  number={13},
  pages={136404},
  year={2008},
  publisher={APS},
  url={https://doi.org/10.1103/PhysRevLett.100.136404}
}

@article{yin2019negative,
  title={Negative flat band magnetism in a spin--orbit-coupled correlated kagome magnet},
  author={Yin, Jia-Xin and Zhang, Songtian S and Chang, Guoqing and Wang, Qi and Tsirkin, Stepan S and Guguchia, Zurab and Lian, Biao and Zhou, Huibin and Jiang, Kun and Belopolski, Ilya and others},
  journal={Nat. Phys.},
  volume={15},
  number={5},
  pages={443--448},
  year={2019},
  publisher={Nature Publishing Group UK London},
  url={https://doi.org/10.1038/s41567-019-0426-7}
}

@article{romer2022superconductivity,
  title={Superconductivity from repulsive interactions on the kagome lattice},
  author={R{\o}mer, Astrid T and Bhattacharyya, Shinibali and Valent{\'\i}, Roser and Christensen, Morten H and Andersen, Brian M},
  journal={Phys. Rev. B},
  volume={106},
  number={17},
  pages={174514},
  year={2022},
  publisher={APS},
  url={https://doi.org/10.1103/PhysRevB.106.174514}
}

@article{yu2012chiral,
  title={Chiral superconducting phase and chiral spin-density-wave phase in a Hubbard model on the kagome lattice},
  author={Yu, Shun-Li and Li, Jian-Xin},
  journal={Phys. Rev. B},
  volume={85},
  number={14},
  pages={144402},
  year={2012},
  publisher={APS},
  url={https://doi.org/10.1103/PhysRevB.85.144402}
}

@article{ferrari2022charge,
  title={Charge density waves in kagome-lattice extended Hubbard models at the van Hove filling},
  author={Ferrari, Francesco and Becca, Federico and Valent{\'\i}, Roser},
  journal={Phys. Rev. B},
  volume={106},
  number={8},
  pages={L081107},
  year={2022},
  publisher={APS},
  url={https://doi.org/10.1103/PhysRevB.106.L081107}
}

@article{profe2024kagome,
  title={Kagome Hubbard model from a functional renormalization group perspective},
  author={Profe, Jonas B and Klebl, Lennart and Grandi, Francesco and Hohmann, Hendrik and D{\"u}rrnagel, Matteo and Schwemmer, Tilman and Thomale, Ronny and Kennes, Dante M},
  journal={Phys. Rev. Res.},
  volume={6},
  number={4},
  pages={043078},
  year={2024},
  publisher={APS},
  url={https://doi.org/10.1103/PhysRevResearch.6.043078}
}

@article{lin2024complex,
  title={Complex magnetic and spatial symmetry breaking from correlations in kagome flat bands},
  author={Lin, Yu-Ping and Liu, Chunxiao and Moore, Joel E},
  journal={Phys. Rev. B},
  volume={110},
  number={4},
  pages={L041121},
  year={2024},
  publisher={APS},
  url={https://doi.org/10.1103/PhysRevB.110.L041121}
}

@article{vidarte2026filling,
  title={Filling-dependent phase competition in the interacting kagome flat band system},
  author={Vidarte, Kevin JU and Cardias, R and Latg{\'e}, A},
  journal={Phys. Rev. B},
  volume={113},
  number={3},
  pages={035134},
  year={2026},
  publisher={APS},
  url={https://doi.org/10.1103/t8nh-qdyl}
}

@article{Wang_2026,
   title={Magnetic fluctuations near the Van Hove singularity in the kagome-lattice Hubbard model at finite doping},
   volume={113},
   ISSN={2469-9969},
   url={http://dx.doi.org/10.1103/yfwy-q6y9},
   DOI={10.1103/yfwy-q6y9},
   number={8},
   journal={Phys. Rev. B},
   publisher={American Physical Society (APS)},
   author={Wang, Jingyao and Jia, Zixuan and Fan, Zenghui and Duan, Qingzhuo and Ma, Tianxing},
   year={2026},
   month=Feb }

@article{mielke1991ferromagnetic,
  title={Ferromagnetic ground states for the Hubbard model on line graphs},
  author={Mielke, Andreas},
  journal={J. Phys. A: Math. Gen.},
  volume={24},
  number={2},
  pages={L73--L77},
  year={1991},
  url={https://iopscience.iop.org/article/10.1088/0305-4470/24/2/005}
}

@article{mielke1991ferromagnetism,
  title={Ferromagnetism in the Hubbard model on line graphs and further considerations},
  author={Mielke, Andreas},
  journal={J. Phys. A: Math. Gen.},
  volume={24},
  number={14},
  pages={3311--3321},
  year={1991},
  url={https://iopscience.iop.org/article/10.1088/0305-4470/24/14/018}
}

@article{kagome_metals_review,
  title = {Kagome metals},
  author = {Di Sante, Domenico and Neupert, Titus and Sangiovanni, Giorgio and Thomale, Ronny and Comin, Riccardo and Checkelsky, Joseph G. and Zeljkovic, Ilija and Wilson, Stephen D.},
  journal = {Rev. Mod. Phys.},
  volume = {98},
  issue = {1},
  pages = {015002},
  numpages = {53},
  year = {2026},
  month = {Feb},
  publisher = {American Physical Society},
  doi = {10.1103/1g9n-wm38},
  url = {https://link.aps.org/doi/10.1103/1g9n-wm38}
}

@article{mielke1992exact,
  title={Exact ground states for the Hubbard model on the Kagome lattice},
  author={Mielke, A},
  journal={J. Phys. A: Math. Gen.},
  volume={25},
  number={16},
  pages={4335--4345},
  year={1992},
  url={https://iopscience.iop.org/article/10.1088/0305-4470/25/16/011}
}

@article{wallerberger2019w2dynamics,
  title={w2dynamics: Local one-and two-particle quantities from dynamical mean field theory},
  author={Wallerberger, Markus and Hausoel, Andreas and Gunacker, Patrik and Kowalski, Alexander and Parragh, Nicolaus and Goth, Florian and Held, Karsten and Sangiovanni, Giorgio},
  journal={Comput. Phys. Commun.},
  volume={235},
  pages={388--399},
  year={2019},
  publisher={Elsevier},
  url={https://doi.org/10.1016/j.cpc.2018.09.007}
}

@article{georges1996dynamical,
  title={Dynamical mean-field theory of strongly correlated fermion systems and the limit of infinite dimensions},
  author={Georges, Antoine and Kotliar, Gabriel and Krauth, Werner and Rozenberg, Marcelo J},
  journal={Rev. Mod. Phys.},
  volume={68},
  number={1},
  pages={13},
  year={1996},
  publisher={APS},
  url={https://doi.org/10.1103/RevModPhys.68.13}
}

@article{gull2011continuous,
  title={Continuous-time Monte Carlo methods for quantum impurity models},
  author={Gull, Emanuel and Millis, Andrew J and Lichtenstein, Alexander I and Rubtsov, Alexey N and Troyer, Matthias and Werner, Philipp},
  journal={Rev. Mod. Phys.},
  volume={83},
  number={2},
  pages={349--404},
  year={2011},
  publisher={APS},
  url={https://doi.org/10.1103/RevModPhys.83.349}
}

@article{metzner1989correlated,
  title={Correlated lattice fermions in $d=\infty$ dimensions},
  author={Metzner, Walter and Vollhardt, Dieter},
  journal={Phys. Rev. Lett.},
  volume={62},
  number={3},
  pages={324},
  year={1989},
  publisher={APS},
  url={https://doi.org/10.1103/PhysRevLett.62.324}
}

@article{kaufmann2023ana_cont,
  title={ana\_cont: Python package for analytic continuation},
  author={Kaufmann, Josef and Held, Karsten},
  journal={Comput. Phys. Commun.},
  volume={282},
  pages={108519},
  year={2023},
  publisher={Elsevier},
  url={https://doi.org/10.1016/j.cpc.2022.108519}
}

\end{document}